\documentstyle{article}
\begin{document}

\title{Electron capture into quantum wells via scattering by
electrons, holes, and optical phonons}
\author{K. K\'alna and M. Mo\v sko \\Institute of Electrical
Engineering, Slovak Academy of Sciences\\ D\'ubravsk\'a cesta 9,
Sk-842 39 Bratislava, Slovakia}
\date{14 November, 1996}
\maketitle
\begin{abstract}
\noindent Electron capture times due to the electron-electron (e-e),
electron-hole (e-h) and elect\-ron-polar optical phonon (e-pop)
interactions are calculated in the $GaAs$ quantum well (QW) with
electron and hole densities $10^{11} cm^{-2}$. The calculated capture
times oscillate as a function of the QW width with the same period but
with different amplitudes. The e-h capture time is two to four orders
larger and the e-e capture time one to three orders larger than the
e-pop capture time. The exceptions are the QW widths near resonance
minima, where the e-e capture time is only $2-3$ times larger and the
e-h capture time $10-100$ times larger. Different physical origin of
the oscillatory behavior is demonstrated for the e-e and e-pop capture
times. Effects of exchange and degeneracy on the e-e capture are
analysed. The exchange effect increases the e-e capture time
approximately two times while the degeneracy does not change the
capture time except for the QW depths and widths near the resonance.
\end{abstract}

\section{Introduction}

The electron capture represents the transition of an electron from the
state above the barrier into a bound state in the quantum well (QW).
In a QW laser the electrons and holes captured in the QW can create
population inversion and participate through mutual recombination in
the laser action. Therefore, the shorter the capture time, the faster
the creation of population inversion, and the laser works at a lower
threshold current and/or with a better high-speed modulation
characteristics.

Since in QW lasers the wave lengths of captured electrons are
comparable with the QW width and their coherence length exceeds the
barrier width \cite{haverkort:pss}, the classical diffusive models
\cite{shichijo,tang} have to be replaced by quantum-mechanical (QM)
ones. Previous QM calculations predicted that the polar optical phonon
(pop) emission induced capture time in the single QW oscillates
\cite{kozyrev:shik,brum:bastard} in dependence on the QW width. These
oscillations were observed in the separate confinement heterostructure
quantum well (SCHQW) \cite{blom:smit}, in the multiple-quantum-well
structure \cite{barros} and in the QW structure with tunnel barriers
\cite{fujiwara}.

The electron and hole capture processes are expected to play an
important role in the optimization of QW laser performance. The hole
capture is faster due to the large hole effective mass and studies of
the capture are therefore focused mainly on electrons
\cite{karol:apl,vurgaftman,blom:haverkort}. The difference between the
electron and hole capture times tends to diminish during the capture
process \cite{ridley} due to an electron-hole attraction and due to
the ambipolar character of the capture. Minimization of the electron
capture time can be achieved mainly by optimizing the QW width and
depth. While the capture due to the electron-polar optical phonon
(e-pop) interaction is dominant outside the oscillation minima, in the
minima the electron-electron (e-e) interaction can play an important
role \cite{karol:apl} and becomes dominant for high enough electron
densities. Increasing influence of the e-e capture with increasing
electron density in the QW was confirmed in Refs.~\ref{allmen:hess}
and~\ref{sotirelis:hess} where the electron degeneracy and the dynamic
screening function in a coupled system of electrons and phonons were
taken into account. Similarly to Ref.~\ref{karol:apl} the exchange
effect was not considered.

The exchange effect was previously included in the carrier-carrier
(c-c) thermalization of photoexcited spin-polarized 2D carriers in the
lowest energy level of the $GaAs$ QW \cite{antonia:prb,mosko:sst}. In
this case the inclusion of the exchange slows down the thermalization
many times. Recently, the exchange has been incorporated into the
intersubband c-c scattering rate in the $GaAs$ QW \cite{mosko:prb}.

In this work we wish to extend our previous calculations of the e-e
interaction induced capture time \cite{karol:apl} by taking into
account the exchange \cite{mosko:prb}. We consider also the electron
capture due to the electron-hole (e-h) interaction where besides the
electron screening also the screening by the holes (often neglected in
similar analyses) is incorporated into the static screening function.
Finally, we analyze different physical origins of the oscillatory
behavior of the e-e and e-pop capture times.

The e-h interaction induced capture time in the SCHQW oscillates with
the same period as the e-e and e-pop capture times but with a larger
oscillation amplitude than the amplitude of the e-e capture time. The
e-h capture time is even larger in the oscillation minima. The
degeneracy influences the value of the e-e capture time only in the
resonance. The exchange effect increases its value two times outside
the resonance and about $10$\% in the resonance.

In section~\ref{rates} the e-h, e-e and e-pop scattering rates are
described including the effects of the exchange and degeneracy in the
e-e interaction. The calculated electron capture times are discussed
in section~\ref{times} and the results are summarized in
section~\ref{concl}.

\section{ Carrier-carrier scattering rate \label{rates} }

Figure~1 shows the band edge profile of the SCHQW structure analyzed
in this work. The structure consists of the $Al_xGa_{1-x}As /GaAs/
Al_xGa_{1-x}As$ QW with $500$-\AA\ $Al_xGa_{1-x}As$ barriers, embedded
between two semiinfinite $AlAs$ layers
\cite{haverkort:pss,blom:haverkort}. The $AlAs$ barrier $V_b$ is
$1.07$ eV. To have the $AlGaAs$ barrier $V_w$ equal to $0.3$ eV we
take the aluminium content $x = 0.305$ \cite{hrivnak}. The carrier
density in the QW is $N_S = 10^{11} cm^{-2}$ to $10^{12} cm^{-2}$ and
the lattice temperature is $8$~K. The same structure has been
considered in previous analyses \cite{haverkort:pss,
blom:smit,karol:apl, blom:haverkort}, since it is of some interest for
optical measurements of the capture time as well as for laser
applications.

The c-c scattering rate in the structure is treated in the Born
approximation according to Ref.~\ref{mosko:prb}. Let the carrier
$\alpha$ occupies the subband $i$ with wave vector ${\bf k}_1$ and the
carrier $\beta$ occupies the subband $j$ with wave vector ${\bf k}_2$.
Due to a mutual Coulomb interaction the carrier $\alpha$ is scattered
into the subband $m$ with wave vector ${\bf k}'_1$ and the partner
carrier $\beta$ is scattered into the subband $n$ with wave vector
${\bf k}'_2$. The c-c scattering rate of a carrier with wave vector
${\bf k}_1$ from the subband $i$ to the subband $j$ can be obtained as
\begin{equation} \lambda^{\alpha \beta}_{im} ({\bf k}_1) =
\frac{1}{N_S A} \sum_{j,n,{\bf k}_2} f^{\beta}_j({\bf k}_2)
\;\lambda^{\alpha \beta}_{ijmn} (g) \quad,\qquad \alpha, \beta = e, h
\quad, \label{lambda} \end{equation}
where $g=|{\bf k}_1-{\bf k}_2|$. The summation over ${\bf k}_2$ is
assumed to include both spin orientations, the summation over $j, n$
involves the subbands below the $AlGaAs$ barrier. $A$ is the
normalization area, $f^{\beta}_j ({\bf k}_2)$ is the Fermi
distribution function of the carriers $\beta$ in the subband $j$, and
$\lambda^{\alpha \beta}_{ijmn} (g)$ is the c-c pair scattering rate.

The electron capture time is reciprocal of the electron capture rate
\begin{equation} \tau^{-1}_{e-\eta} = \frac{ \sum_{i,m,{\bf k}_1}
f^e_i ({\bf k}_1) \; \lambda^{e \eta}_{im} ({\bf k}_1) }{
\sum_{i,{\bf k}_1} f^e_i ({\bf k}_1)} \quad,\qquad \eta=e,h.
\label{captime} \end{equation}
where the summation over $i$ ($m$) includes only the subbands above
(below) the $AlGaAs$ barrier and $f^e_i ({\bf k}_1)$ is the
electron distribution in the subband $i$.

\subsection{Electron-hole pair scattering rate}

The e-h pair scattering rate is calculated considering the screening
by electrons and holes occupying the lowest energy subband (see
Appendix). It reads
\begin{eqnarray} \lefteqn{ \lambda^{eh}_{ijmn} (g) = \frac{N_S m_r
e^4}{16\pi \hbar^3 \kappa^2} \int_0^{2\pi} d\theta\:} \nonumber \\ & &
\times {\left[ F^{eh}_{ijmn} (q) - \frac{q^e_s}{q \epsilon^{eh} (q)}
\; F^{ee}_{i1m1} (q) G_{1j1n} (q) + \frac{q^h_s}{q \epsilon^{eh} (q)}
\; F^{eh}_{i1m1} (q) H_{1j1n} (q) \right]}^2 q^{-2} \quad,
\label{paireh}
\end{eqnarray}
where the e-h static screening function $\epsilon^{eh} (q)$ is given
as
\begin{eqnarray} \epsilon^{eh} (q) &=& \left( 1+ \frac{q^e_s}{q}
F^{ee}_{1111} (q) \right) \left( 1+ \frac{q^h_s}{q} F^{hh}_{1111} (q)
\right) - \frac{q^e_s}{q} F^{eh}_{1111} (q) \frac{q^h_s}{q}
F^{he}_{1111} (q) \quad, \label{screef} \\ G_{1j1n} (q) &=&
F^{eh}_{1j1n} (q) \left[ 1+ \frac{q^h_s}{q} F^{hh}_{1111} (q) \right]
- \frac{q^h_s}{q} F^{hh}_{1j1n} (q) F^{he}_{1111} (q) \quad ,
\nonumber \\ H_{1j1n} (q) &=& F^{hh}_{1j1n} (q) \left[ 1+
\frac{q^e_s}{q} F^{ee}_{1111} (q) \right] - \frac{q^e_s}{q}
F^{eh}_{1j1n} (q) F^{eh}_{1111} (q) \quad , \nonumber \\ q &=&
\frac{1}{2} {\left[ 2g^2 + \frac{4m_r}{\hbar^2} E^{eh}_S - 2g {
\left(g^2 + \frac{4m_r}{\hbar^2} E^{eh}_S \right)}^{1/2} \cos \theta
\right]}^{1/2} \quad, \label{qeh} \end{eqnarray}
and $E^{eh}_S = E^e_i + E^h_j - E^e_m - E^h_n$. In the above formulae
the reduced effective mass $m_r = 2 m_e m_h / ( m_e + m_h )$, the
relative vector ${\bf g} = m_r ({\bf k}_2/m_h -
{\bf k}_1/m_e)$, $\kappa$ is the static permittivity, the electron
and hole effective masses are denoted as $m_e$ and $m_h$, and the
electron and hole subband energies as $E^e_{\gamma} \;(\gamma=i,m)$
and $E^h_{\delta} \;(\delta=j,n)$, respectively. Finally, $q^e_s = e^2
m_e / (2 \pi \kappa \hbar^2) \:f^e_1 ({\bf k}_2=0)$ and $q^h_s =
e^2 m_h / (2 \pi \kappa \hbar^2) \:f^h_1 ({\bf k}_2=0)$.

The form factors in Eq.~(\ref{paireh}) are defined as
\begin{equation} F^{\alpha \beta}_{ijmn} (q) = \int^{\infty}_{-\infty}
dz_1 \; \int^{\infty}_{-\infty} dz_2\; \chi^{\alpha}_i
(z_1)\:\chi^{\beta}_j (z_2) \;e^{-q|z_1-z_2|} \;\chi^{\alpha}_m (z_1)
\:\chi^{\beta}_n (z_2) \quad, \qquad \alpha,\beta = e,h.
\label{formfac} \end{equation}
where the wave function $\chi^{\alpha}_{\gamma}$ of the carrier
$\alpha$ in the subband $\gamma \;(\gamma=i,j,m,n)$ is obtained
assuming the $x$-dependent carrier effective mass and the flat
$\Gamma$-band with parabolic energy dispersion, both properly
interpolated between $GaAs$ and $AlAs$ \cite{hrivnak}.

\subsection{Electron-electron pair scattering rate}

The e-e pair scattering rate is found in a similar way (see Appendix)
as the e-h pair scattering rate. Taking into account only the
screening by electrons in the lowest subband, one gets
\begin{equation} \lambda^{ee}_{ijmn} (g) = \frac{N_S m_e e^4}{16\pi
\hbar^3 \kappa^2} \int_0^{2\pi} d\theta\: {\left[ F^{ee}_{ijmn} (q) -
\frac{q^e_s}{q \epsilon^{e} (q)} F^{ee}_{i1m1} (q) F^{ee}_{1j1n} (q)
\right]}^2 q^{-2} \quad, \label{pairee} \end{equation}
where the static screening function $\epsilon^{e} (q)$ reads
\begin{eqnarray} \epsilon^{e} (q) &=& 1+(q^e_S/q)
F^{ee}_{1111}(q)\;f^e_1 ({\bf k}_1=0) \quad, \label{statsc} \\ q
&=& \frac{1}{2} {\left[ 2g^2 + \frac{4m_e}{\hbar^2} E^e_S - 2g {
\left(g^2 + \frac{4m_e}{\hbar^2} E^e_S \right)}^{1/2} \cos \theta
\right]}^{1/2}, \label{qee} \end{eqnarray}
and $E^e_S = E^e_i + E^e_j - E^e_m - E^e_n$. In Eq.~(\ref{statsc}) the
static screening by the holes is omitted assuming that the holes are
too slow (due to their large effective masses) to follow the fast
changes of electron positions \cite{antonia:prb}. This is a
so-called quasi-dynamic screening model.

It is worth mentioning that the screening [i.e. the term containing
the screening function~$\epsilon^{e}~(q)$] disappears in
Eq.~(\ref{pairee}) for those transitions in which $F^{ee}_{i1m1} (q)
= 0$ and/or $F^{ee}_{1j1n} (q) = 0$. This happens when $\chi^e_i
(z_1)$ is symmetric (antisymmetric) and $\chi^e_m (z_1)$ is
antisymmetric (symmetric), and/or when the same holds for $\chi^e_j
(z_2)$ and $\chi^e_n (z_2)$.

For example, when the QW contains two energy subbands, the screening
effect disappears for the transitions $i, 1 \to 1, 2
\;(i=3,5,\ldots)$. Then the e-e pair scattering rate reads
\begin{equation} \lambda^{ee}_{i112} (g) = \frac{N_S m_e e^4}{16\pi
\hbar^3 \kappa^2} \int_0^{2\pi} d\theta\: \frac{ {\left( F^{ee}_{i112}
(q) \right)}^2}{q^2} \quad. \label{pairun} \end{equation}
Note that the same effect appears also in the e-h pair scattering rate
(\ref{paireh}).

After simple manipulations, the integrand in Eq.~(\ref{pairee}) can be
simplified assuming
\begin{equation} \frac{F^{ee}_{ijmn} (q) F^{ee}_{1111}
(q)}{F^{ee}_{i1m1} (q) F^{ee}_{1j1n} (q)} \approx 1 \label{ffapp}
\end{equation}
into the form \cite{mosko:prb}
\begin{equation} \lambda^{ee}_{ijmn} (g) = \frac{N_S m_e e^4}{16\pi
\hbar^3 \kappa^2} \int_0^{2\pi} d\theta\: \frac{{|F^{ee}_{ijmn}
(q)|}^2}{q^2 \;\epsilon^{e} (q)^2} \quad. \label{pairap}
\end{equation}

To justify the approximation (\ref{ffapp}), the e-e capture times
calculated for various screening functions are compared in Fig.~2. In
this calculation the electron distribution function $f^e_i ({\bf
k}_1)$ in the formula (\ref{captime}) was taken as a constant
distribution up to the pop energy above the $AlGaAs$ barrier, which
roughly models the injected barrier distribution after a rapid phonon
cooling \cite{brum:bastard}. The capture time obtained from the pair
scattering rate (\ref{pairap}) with the screening function
(\ref{statsc}) is almost the same as the capture time obtained from
the pair scattering rate (\ref{pairee}). If (following
Ref.~\ref{gl:apl}) we use Eq.~(\ref{pairap}) with $F^{ee}_{1111} (q) =
1$ in the screening function (\ref{statsc}), the calculated capture
time overestimates both capture times more than by $20$\%.

\subsection{Electron-electron pair scattering rate with degeneracy and
exchange}

To include the Pauli exclusion principle in the e-e scattering rate
(\ref{pairee}) we can start from the equation
\begin{eqnarray} \lambda^{ee}_{ijmn} (g) &=& \frac{N_S m_e e^4}{8\pi
\hbar^3 \kappa^2} \int \;d{\bf k}'_2 \left[ 1 - f^e_m({\bf k}_1+{\bf
k}_2-{\bf k}'_2) \right] \left[ 1 - f^e_n({\bf k}'_2) \right]
\nonumber \\ &\times& \frac{{ \left( F^{ee}_{ijmn} (q_l) \right) }^2}
{q_l^2 \epsilon^{e} (q_l)^2}\; \delta \!\left[ \frac{\hbar^2}{2m_e}
({{\bf k}_1}^2 +{{\bf k}_2}^2 -{{\bf k}'_1}^2 -{{\bf k}'_2}^2 ) +
E^e_S \right] \quad, \label{source} \end{eqnarray}
in which ${\bf k}'_1={\bf k}_1 +{\bf k}_2 -{\bf k}'_2$. The delta
function simplifies the integration over ${\bf k}'_2$ in
Eq.~(\ref{source}) and then the e-e pair scattering rate with
degeneracy is given by
\begin{equation} \lambda^{ee}_{ijmn} (g) = \frac{N_S m_e e^4}{8\pi
\hbar^3 \kappa^2} \sum_{l=1}^2 \int_0^{2\pi} d\theta\: \frac{p_l}{|D
-2p_l|} \left[ 1 - f^e_m(r_l) \right] \left[ 1 -
f^e_n(p_l) \right] \frac{{ \left( F^{ee}_{ijmn} (q_l) \right)
}^2}{q_l^2 \epsilon^{e} (q_l)^2} \label{pairde} \quad, \end{equation}
where
\begin{eqnarray*} q_l &=& {[{p_l}^2 + k_2^2 - 2k_2 p_l
\cos(\theta-\phi)]}^{1/2} \quad, \qquad l=1,2 \\ r_l &=& {[k_1^2 +
k_2^2 + p_l -2k_1 k_2 \cos \phi - 2k_1p_l \cos \theta + 2k_2 p_l \cos
(\theta - \phi) ]}^{1/2} \quad, \\ p_1 &=& \frac{1}{2}D+\frac{1}{2}
{\left[ D^2 + \frac{4m}{\hbar^2}E^e_S - 4k_1k_2 \cos\phi
\right]}^{1/2} \quad, \\ p_2 &=& \frac{1}{2} D-\frac{1}{2} {\left[ D^2
+ \frac{4m}{\hbar^2}E^e_S - 4k_1k_2 \cos\phi \right]}^{1/2} \quad, \\
D &=& k_1 \cos\theta + k_2 \cos(\theta-\phi) \quad. \end{eqnarray*}
The angles $\theta$ and $\phi$ in the above expressions are between
the wave vectors ${\bf k}_1, {\bf k}'_2$ and ${\bf k}_1, {\bf k}_2$;
respectively.

Electrons are undistinguishable particles. Therefore, the scattering
rate (\ref{pairee}) should include the exchange effect
\cite{antonia:prb}. According to Ref.~\ref{mosko:prb} the exchange can
be incorporated into the intersubband e-e scattering rate
(\ref{pairde}) using the replacement
\begin{equation} \frac{{|F^{ee}_{ijmn}(q_l)|}^2}{q_l^2 \epsilon^{e}
(q_l)^2} \mapsto \frac{1}{2} \left[
\frac{{|F^{ee}_{ijmn}(q_l)|}^2}{q_l^2 \epsilon^{e} (q_l)^2} +
\frac{{|F^{ee}_{ijnm} (q'_l)|}^2}{{q'_l}^2 \epsilon^{e} (q'_l)^2} -
\frac{F^{ee}_{ijmn} (q_l) F^{ee}_{ijnm} (q'_l)} {q_l \epsilon^{e}
(q_l) q'_l \epsilon^{e} (q'_l)} \right] \quad, \label{exch}
\end{equation}
where
\begin{displaymath} q'_l = {[{p_l}^2 + k_1^2 - 2k_1 p_l
\cos(\theta)]}^{1/2} \quad, \qquad l=1,2. \end{displaymath}

\subsection{Electron-polar optical phonon scattering rate}

The e-pop scattering rate of an electron with wave vector
${\bf k}_1$ from subband $i$ to subband $m$ for a spontaneous phonon
emission only reads \cite{goodnick:lugli,mosk:thesis}
\begin{eqnarray} \lambda^{epop}_{im} ({\bf k}_1) = \frac{e^2 \omega
m_e}{8 \pi \hbar^2} \left( \frac{1}{\kappa_{\infty}} -
\frac{1}{\kappa} \right) \int_0^{2 \pi} d\theta\; \frac{F^{ee}_{iimm}
(q)}{q \;\epsilon^{e} (q) } \quad, \label{lampop} \\ q = {\left[
2k_1^2 + \frac{2m_e}{\hbar^2} E^P_S - 2k_1 { \left(k_1^2 +
\frac{2m_e}{\hbar^2} E^P_S \right)}^{1/2} \cos \theta \right]}^{1/2}
\quad, \label{qpop} \end{eqnarray}
where $E^P_S = E_i-E_m-\hbar\omega$,\ $\hbar\omega$ is the pop energy
and $\kappa_{\infty}$ is the high frequency permittivity. To obtain
the e-pop capture time the same formula as (\ref{captime}) can be
used after replacing $\eta$ by $pop$.

\section{ Electron capture times \label{times} }

Figure~3 shows the electron capture time versus the QW width for e-e,
e-h, and e-pop interactions where, in the case of the e-e interaction,
the results obtained with and without exchange effect are
distinguished. In these calculations the function $f^e_i ({\bf k}_1)$
in the formula (\ref{captime}) is taken as a constant distribution up
to the pop energy above the $AlGaAs$ barrier. The capture times
oscillate with the QW width and reach the minimum whenever a new bound
state merges into the QW \cite{karol:apl}. At small QW widths, when
the QW contains only one bound state, the e-h and e-e capture times
increase with an increasing QW width. At the QW widths at which the
second bound state merges into the QW ($\sim 30$ \AA\ for holes, $46$
\AA\ for electrons), the e-h and e-e capture times decrease suddenly
by several orders of magnitude. With a further increase of the QW
width the oscillatory behavior persists, but the oscillations become
smooth. To understand this feature we split in Fig.~4(a) the total e-e
capture time from Fig.~3 into the e-e capture time to states $1, 1$
(full squares) and the e-e capture time to states $1, 2$ and $2, 1$
(full triangles). A smooth decrease of the total capture time from its
maximum value at $w=73$ \AA\ is due to the decrease of the capture
time to states $1, 1$ with $w$. This decrease is caused by the
increase of relevant form factors [Fig.~4(c)] with $w$. At the same
time the capture time to state $1, 2$ and $2, 1$ increases with $w$
due to the decrease of relevant form factors [Fig.~4(b)] with $w$.

The behaviour of the e-pop capture time is quite different. The
decrease of the e-pop capture time to its oscillation minima is smooth
even when the first minimum appears. The e-pop capture time curve does
not show a resonant drop for the QW widths near the resonances,
because the barrier electrons occupy the states below the threshold
for the pop emission and cannot be scattered into the subband which is
in the resonance with the top of the QW. A further increase of the QW
width shifts the resonant subband deeper into the QW and the e-pop
scattering into this subband smoothly increases. The exception is a
monoenergetic distribution with the energy close to the pop energy. In
such case the e-pop scattering into the resonance subband is not
prohibited and a resonant decrease of the e-pop capture time takes
place \cite{sotirelis:hess}. Similarly to the e-e interaction case we
split in Fig.~5(a) the total e-pop capture time from Fig.~3 into the
e-pop capture time to states $1, 1$ (open squares) and the e-pop
capture time to states $2, 2$ (open triangles). Transitions to the
highest subband in the QW play always a more important role than
transitions to the lower subbands. This fact explains the behaviour of
the relevant form factors in dependence on the QW width in Figs.~5(b)
and~5(c).

The total electron capture time which is, in fact, measured in an
experiment can be obtained as
\begin{equation} \tau_{e-tot} = \frac{\tau_{e-e} \tau_{e-pop}}{
\tau_{e-e} + \tau_{e-pop} } \quad. \label{totime} \end{equation}
In Fig.~6 the e-pop capture time (full circles) is compared with the
total electron capture times (open symbols) obtained using the e-e
capture time with exchange for the electron density $N_S=10^{11}
cm^{-2}$ and $N_S=10^{12} cm^{-2}$. This figure depicts how the e-e
interaction affects the total electron capture time near the resonance
and how its effect decreases the total electron capture time when the
electron density is higher. A direct comparison of our total electron
capture time with the experiments \cite{blom:smit,barros,fujiwara} is
not possible because we use the step distribution function for an
active laser regime. Nevertheless, we have obtained the resonances in
the capture time oscillations by the same QW width as resonances
measured in the experiments. The direct comparison requires to take
into account the ambipolar capture time which can only be calculated
when the hole capture time is fitted to the experimental data
\cite{blom:smit,ridley}.

To directly detect the e-e capture time it is necessary to suppress
the e-pop interaction. A proper structure for the e-pop interaction
suppression is the structure with the QW depth smaller than the pop
energy. If in the time-resolved optical experiment \cite{blom:smit}
such structure would be irradiated by a short laser pulse, the excited
carriers above the barrier would thermalize to the Boltzmann
distribution with the electron temperature $T_e$ during several
picoseconds \cite{blom:smit,snoke} and after that they would be
captured via the e-e interaction into the QW. Figure~7 shows the
electron capture time versus the QW depth for the QW width equal to
46~\AA. Here we assume that the QW depth varies with the aluminium
content $x$ according to the relation $V_w = (0.9456 x + 0.1288 x^2)$
eV \cite{hrivnak}. In this calculation the distribution function
$f^e_i ({\bf k}_1)$ in the capture time formula (\ref{captime}) is
taken as the Boltzmann distribution function at the electron
temperature $T_e = 70 K$. This choice corresponds to the assumption
that the electrons are optically excited only a few meV above the
$AlGaAs$ barrier. Figure~7 also shows that the e-e capture dominates
if the QW depth is less than $0.04$ eV. The degeneracy and exchange
affect the e-e capture time substantially at a small QW depth.

The electron capture times from Fig.~8 illustrate the inclusion and
exclusion of the degeneracy and exchange effect in the e-e interaction
with an increasing electron density for the shallow QW ($V_w = 0.05
eV, w = 46$ \AA). In such structures, recently also considered in
Ref.~\ref{preisel}, the e-e capture time is of similar importance as
the e-pop capture time for higher electron densities although it was
calculated including the degeneracy and exchange effect in the e-e
interaction which increases the capture time by about two times.

Outside the resonance the exchange increases the e-e capture time by
about two times due to the following reasons. In case that both
electrons are scattered into the same subband ($m=n$), all three terms
in the substitution (\ref{exch}) have almost the same magnitude (which
we verify numerically) because the values of $q$ and $q'$ are very
close. In case that both electrons are scattered into different
subbands ($m \neq n$), it is necessary to take into account that the
substitution (\ref{exch}) is summed in the formulae (\ref{lambda}) and
(\ref{captime}) over the final states $m$ and $n$. In the sum the
terms with final states $m, n$ and $n, m$ can be rewritten as
\begin{eqnarray} \lefteqn{\frac{1}{2} \left[
\frac{{|F_{ijmn}(q)|}^2}{q^2 \epsilon (q)^2} +
\frac{{|F_{ijnm}(q)|}^2}{q^2 \epsilon (q)^2} \right] + \frac{1}{2}
\left[ \frac{{|F_{ijnm} (q')|}^2}{{q'}^2 \epsilon (q')^2} +
\frac{{|F_{ijmn} (q')|}^2}{{q'}^2 \epsilon (q')^2} \right]} \nonumber
\\ & & -\frac{1}{2} \left[ \frac{F_{ijmn} (q) F_{ijnm} (q')} {q
\epsilon (q) q' \epsilon (q')} + \frac{F_{ijnm} (q) F_{ijmn} (q')} {q
\epsilon (q) q' \epsilon (q')} \right] \label{sumex} \quad.
\end{eqnarray}
All three terms in Eq.~(\ref{sumex}) have almost the same magnitude
which we verify numerically. Consequently, the presence of the third
(interference) term decreases the value of the expression
(\ref{sumex}) two times in comparison with the case when the exchange
effect (i. e., the interference term) is neglected. Finally, in the
resonance the exchange increases the capture time only by $10$\%. This
is due to the fact that in the resonance the values of $q$ and $q'$
strongly differ and the interference term becomes much smaller than
direct terms.

The effect of degeneracy on the capture time is negligible when the
difference between the lowest energy subband above the barrier and the
highest energy subband in the QW is large compared to the quasi-Fermi
energy of the electrons in the QW. In such case most of final states
in the QW are unoccupied and the e-e capture times with and without
degeneracy are quite close. Near the resonance, when the highest
energy subband in the QW is close to the lowest subband above the
barrier, the degeneracy strongly reduces the number of available final
states despite the fact that the highest subband is essentially
unoccupied. This is due to the fact that the electron captured in the
highest subband of the QW can exchange only a small amount of energy
with the scattering partner in the lowest subband of the QW. The
available final states of the scattering partner are therefore blocked
by the $8$-K Fermi distribution in the QW.

\section{ Conclusions \label{concl} }

The e-h, e-e and e-pop capture times have been calculated in the SCHQW
for the carrier density $10^{11} cm^{-2}$ and $10^{12} cm^{-2}$. All
three capture times oscillate as a function of the QW width with the
same period but with very different amplitudes. The e-h capture time
is not only much greater than the e-pop capture time but also greater
than the e-e capture time even for the QW widths near the resonance.
In the resonance the e-e interaction plays a role together with the
e-pop interaction and improves the capture efficiency of the QW
\cite{karol:apl}. Since the increase of the electron density decreases
the e-e capture time, it is expected that the e-h capture time will
decrease similarly. Therefore, the influence of the e-h interaction
should be considered only for high density and in resonances.

We find that the electron capture time oscillates as a function of the
depth and reaches the oscillation minimum when a new bound state
merges into the QW. At the same time the effect of the degeneracy and
exchange effect on the e-e capture time has been studied. The
degeneracy increases the capture time approximately $50$\% only in the
resonant minima. The inclusion of the exchange effect into the e-e
interaction increases the e-e capture time by about two times. To
illustrate the quantitative importance of the exchange effect found in
this paper, we mention that the quantitative changes due to the
dynamic screening and coupling between electrons and phonons (not
considered in this work) are much smaller \cite{allmen:hess}.
Therefore, the exchange effect should be considered in the e-e
interaction induced capture.

\section{Acknowledgment}

Numerous useful discussions with A. Mo\v skov\'a are greatly
appreciated. We also thank her and B. Olejn\'\i kov\'a for a careful
reading of the manuscript. This work was supported by the Slovak Grant
Agency for Science under contract No.~$2/1194/94$.

\appendix \section*{Appendix: Carrier-carrier interaction
with multisubband static screening}
%%%%%%%%%%%%%%%%%%%%%%%%%%%%%%%%%%%%%%%%%%%%%%%%%%%%%%%%%%%%%%%%%%%%%
%      Definition of the new counter in appendix                    %
%%%%%%%%%%%%%%%%%%%%%%%%%%%%%%%%%%%%%%%%%%%%%%%%%%%%%%%%%%%%%%%%%%%%%
\setcounter{equation}{0}
\renewcommand{\theequation}{A\arabic{equation}}

The statically screened intercarrier interaction in the multisubband
2D system, $V ({\bf Q},z_1,z_2)$, can be obtained by solving the
integral Poisson equation \cite{mosk:thesis,yokoyama:hess}
\begin{eqnarray} \lefteqn{ V ({\bf Q},z_1,z_2) = \frac{e}{2\kappa Q}
e^{-Q|z_1-z_2|} -} \nonumber \\ &-& \sum^L_{l=1} \frac{Q^e_l}{Q}
\int^{\infty}_{-\infty} dz''\; {\left( \phi^e_l (z'') \right)}^2
e^{-Q|z_1-z''|} \int^{\infty}_{-\infty} dz'\; {\left( \phi^e_l (z')
\right)}^2 V ({\bf Q},z',z_2) - \nonumber \\ &-& \sum^K_{k=1}
\frac{Q^h_k}{Q} \int^{\infty}_{-\infty} dz''\; {\left( \phi^h_l (z'')
\right)}^2 e^{-Q|z_1-z''|} \int^{\infty}_{-\infty} dz'\; {\left(
\phi^h_l (z') \right)}^2 V ({\bf Q},z',z_2) \quad, \label{fehcp}
\end{eqnarray}
where $L (K)$ is the number of electron (hole) subbands, the first
term on the right hand side is the bar Coulomb interaction, the second
and third terms describe the screening by electrons and holes,
respectively, and
\begin{displaymath} Q^e_l = \frac{e^2 m_e}{2 \pi \kappa \hbar^2} f^e_l
({\bf k}=0) \quad,\qquad Q^h_k = \frac{e^2 m_h}{2 \pi \kappa
\hbar^2} f^h_k ({\bf k}=0) \end{displaymath}
are the 2D electron and 2D hole screening constants in the $l$-th
electron and $k$-th hole subband, respectively.

If the Coulomb potential (\ref{fehcp}) is multiplied by the electron
wave function $\phi^e_l$ for $l=1,2,\ldots,L$ and by the hole wave
function $\phi^h_k$ for $k=1,2,\ldots,K$, and then integrated over
$z'$, the following two sets of mutually coupled equations are found
\begin{eqnarray} \left( 1+A^e_{pp} \right) x^e_p + \sum^L_{ l=1, l\neq
p} C^{ee}_{pl} x^e_l + \sum^K_{k=1} C^{he}_{pk} x^h_k = B^e_p \quad,
\qquad p=1,2,\ldots,L \quad, \nonumber \\ \sum^L_{l=1} C^{eh}_{rl}
x^e_l + \left( 1+A^h_{rr} \right) x^h_r + \sum^K_{k=1, k\neq r}
C^{hh}_{rk} x^h_k = B^h_r \quad, \qquad r=1,2,\ldots,K \quad,
\label{ehleq} \end{eqnarray}
where
\begin{eqnarray*} x^{\eta}_i &=& \int^{\infty}_{-\infty} dz'\; {\left(
\phi^{\eta}_i (z') \right)}^2 V(Q,z',z_2) \quad, \\ A^{\eta}_{ii} &=&
\frac{Q^{\eta}_i}{Q} \int^{\infty}_{-\infty} dz'\; {\left(
\phi^{\eta}_i (z') \right)}^2 \int^{\infty}_{-\infty} dz''\; {\left(
\phi^{\eta}_i (z'') \right)}^2 e^{-Q|z'-z''|} \quad, \\ B^{\eta}_i &=&
\frac{e}{2 \kappa Q} \int^{\infty}_{-\infty} dz'\; {\left(
\phi^{\eta}_i (z') \right)}^2 e^{-Q|z'-z_2|} \quad; \qquad \eta = e,h
\quad, \\ C^{\eta \vartheta}_{ij} &=& \frac{Q^{\eta}_j}{Q}
\int^{\infty}_{-\infty} dz'\; {\left( \phi^{\eta}_i (z') \right)}^2
\int^{\infty}_{-\infty} dz''\; {\left( \phi^{\vartheta}_j (z'')
\right)}^2 e^{-Q|z'-z''|} \quad; \qquad \eta, \vartheta = e,h \quad.
\end{eqnarray*}

Unlike the integral equation (\ref{fehcp}) the linear equations
(\ref{ehleq}) can be easily solved. When the solutions $x^{\eta}_i$
are set back into the Eq.~(\ref{fehcp}), one obtains the intercarrier
interaction $V ({\bf Q},z_1,z_2)$ in a closed form
\begin{eqnarray} V ({\bf Q},z_1,z_2) &=& \frac{e}{2\kappa Q}
e^{-Q|z_1-z_2|} - \sum^L_{l=1} \frac{Q^e_l}{Q} \int^{\infty}_{-\infty}
dz''\; {\left( \phi^e_l (z'') \right)}^2 e^{-Q|z_1-z''|}
\frac{{\cal D}_l (z_2)}{\cal D} - \nonumber \\ &-& \sum^K_{k=1}
\frac{Q^h_k}{Q} \int^{\infty}_{-\infty} dz''\; {\left( \phi^h_k (z'')
\right)}^2 e^{-Q|z_1-z''|} \frac{{\cal D}_{k+L} (z_2)}{\cal D}
\quad, \label{ehmcoul} \end{eqnarray}
where ${\cal D}_i/{\cal D} = x^{\eta}_i$ is the $i$-th solution of the
system (\ref{fehcp}) [${\cal D}_i$ and ${\cal D}$ are appropriate
determinants]. The Coulomb matrix element
\begin{equation} \int^{\infty}_{-\infty} dz_1\;
\int^{\infty}_{-\infty} dz_2\; \phi_i (z_1) \phi_j (z_2) V
({\bf Q},z_1,z_2) \phi_m (z_1) \phi_n (z_2) \label{matehmc}
\end{equation}
is then easy to evaluate.

The e-h scattering rate (\ref{paireh}) and the e-e scattering rate
(\ref{pairee}) are obtained using the Coulomb matrix element
(\ref{matehmc}) for $L=K=1$ and for $L=1, K=0$, respectively. Taking
$K=0$ in the e-e interaction we neglect the screening by the holes.
This corresponds to the so-called quasi-dynamic approximation
\cite{mosko:prb}, in which heavy holes are not able to follow the fast
changes of electron positions. Finally, as in previous papers
\cite{blom:haverkort,sotirelis:hess}, we restrict ourselves to the
screening by quasi-equilibrium carriers in the QW. In our conditions
(the density of the carriers equal to $10^{11} cm^{-2}$ or $10^{12}
cm^{-2}$, the lattice temperature $8$~K) the quasi-equilibrium
carriers occupy mainly the lowest subband of the QW. Therefore, we
neglect $L$ and $K$ greater than $1$.

\newpage
{\large Figure captions}
\medskip

{\bf Fig. 1} Conduction band edge diagram (schematic) and geometry of
the separate confinement heterostructure quantum well considered in
our calculations.
\medskip

{\bf Fig. 2} E-e capture time as a function of QW width for various
static screening models. The capture time calculated using the e-e
scattering rate (\ref{pairee}) [open triangles] is compared with the
capture times calculated using the e-e scattering rate (\ref{pairap})
screened by the screening function (\ref{statsc}) with realistic
$F_{1111}(q)$ [full circles] and with $F_{1111}(q) = 1$ [full
squares].
\medskip

{\bf Fig. 3} Electron capture time versus the QW width for the e-pop
interaction (full circles), the e-h interaction (open triangles), the
e-e interaction without exchange effect (open squares), and the e-e
interaction with exchange effect (crosses).
\medskip

{\bf Fig. 4} (a) E-e capture time versus the QW width. The total
capture time from Fig.~3 (open circles) is split into the e-e capture
time to state $1, 1$ (full squares) and into the e-e capture time to
states $1, 2$ and $2, 1$ (full triangles). (b) Squares of the e-e form
factors $F_{i112}$ and $F_{i121}$ versus the QW width for $q=3\times
10^8 m^{-1}$. (c) Squares of the e-e form factors $F_{i111}$ versus
the QW width for $q=3\times 10^8 m^{-1}$.
\medskip

{\bf Fig. 5} (a) E-pop capture time versus the QW width. The total
capture time from Fig.~3 (full circles) is split into the e-pop
capture time to state $1, 1$ (open squares) and into the e-pop capture
time to state $2, 2$ (open triangles). (b) The e-pop form factors
$F_{ii22}$ versus the QW width for $q=3\times 10^8 m^{-1}$. (c) The
e-pop form factors $F_{ii11}$ versus the QW width for $q=3\times 10^8
m^{-1}$.
\medskip

{\bf Fig. 6} Total electron capture times versus the QW width for the
electron density $10^{11} cm^{-2}$ (open triangles) and for the
electron density $10^{12} cm^{-2}$ (open circles) are compared with
the e-pop capture time (full circles). Electron capture times for the
e-e interaction with exchange effect are also shown for the electron
density $10^{11} cm^{-2}$ (crosses) and $10^{12} cm^{-2}$ (plusses).
\medskip

{\bf Fig. 7} Electron capture time as a function of QW depth for
the QW width of $46$\AA. Open circles are for the e-e interaction
without degeneracy and exchange effect, open squares are for the e-e
interaction with degeneracy, open triangles are for the e-e
interaction with degeneracy and exchange effect, and full circles are
for the e-pop interaction induced capture time. In the calculations
the distribution function $f^e_i ({\bf k}_1)$ in the capture time
formula (\ref{captime}) is the Boltzmann distribution at the electron
temperature $70$~K.
\medskip

{\bf Fig. 8} Electron capture time versus the electron density for the
QW with $V_w=0.05 eV$ and $w=46$~\AA. Open circles (triangles)
represent the capture time for the e-e interaction without (with)
degeneracy and exchange effect, respectively, and full circles
represent the capture time for the e-pop interaction.


\begin{thebibliography}{99}
\bibitem{haverkort:pss} J. E. M. Haverkort, P. W. M. Blom, P. J. Van
Hall, J. Claes, and J. H. Wolter, Phys. Status Solidi {\bf B~188},
139 (1995).
\bibitem{shichijo} H. Shichijo, R. M. Kolbas, N. Holonyak, Jr., R. D.
Dupuis and P. D. Dapkus, Solid State Commun. {\bf 27}, 1029 (1978).
\bibitem{tang} J. Y. Tang, K. Hess, N. Holonyak, Jr., J. J. Coleman,
and P. D. Dapkus, J. Appl. Phys. {\bf 53}, 6083 (1982).
\bibitem{kozyrev:shik} S. V. Kozyrev and A. Ya. Shik, Fiz. Tech.
Poluprovodn. {\bf 19}, 1667 (1985) [Sov. Phys. Semicond.
{\bf 19}, 1024 (1985)].
\bibitem{brum:bastard} J. A. Brum, and G. Bastard, Phys. Rev.
{\bf B~33}, 1420 (1986).
\bibitem{blom:smit} P. W. M. Blom, C. Smit, J. E. M. Haverkort, and J.
H. Wolter, Phys. Rev. {\bf B~47}, 2072 (1993).
\bibitem{barros} M. R. X. Barros, P. C. Becker, S. D. Morris, B.
Deveaud, A. Regreny, and F. Beisser, Phys. Rev. {\bf B 47}, 10951
(1993).
\bibitem{fujiwara} A. Fujiwara, Y. Takahashi, S. Fukatsu, Y. Shiraki,
and R. Ito, Phys. Rev. {\bf B 51}, 2291 (1995).
\bibitem{karol:apl} \label{karol:apl} K. K\'alna, M. Mo\v sko, and F.
M. Peeters, Appl. Phys. Lett. {\bf 68}, 117 (1996).
\bibitem{vurgaftman} I. Vurgaftman, Y. Lam, and J. Singh, Phys. Rev.
{\bf B~50}, 14309 (1994).
\bibitem{blom:haverkort} P. W. M. Blom, J. E. M. Haverkort, P. J. van
Hall, and J. H. Wolter, Appl. Phys. Lett. {\bf 62}, 1490 (1993).
\bibitem{ridley} B. K. Ridley, Phys. Rev. {\bf B 50}, 1717 (1994).
\bibitem{allmen:hess} \label{allmen:hess} P. Sotirelis, P. von Allmen,
and K. Hess, Phys. Rev. {\bf B 47}, 12 744 (1993).
\bibitem{sotirelis:hess} \label{sotirelis:hess} P. Sotirelis and K.
Hess, Phys. Rev. {\bf B 49}, 7543 (1994).
\bibitem{antonia:prb} A. Mo\v skov\'a and M. Mo\v sko, Phys. Rev.
{\bf B~49}, 7443 (1994).
\bibitem{mosko:sst} M. Mo\v sko and A. Mo\v skov\'a, Semicond. Sci.
Technol. {\bf 9}, 478 (1994).
\bibitem{mosko:prb} \label{mosko:prb} M. Mo\v sko, A. Mo\v skov\'a,
and V. Cambel, Phys. Rev. {\bf B~51}, 16 860 (1995).
\bibitem{hrivnak} \v L. Hrivn\'ak, Appl. Phys. Lett. {\bf 56}, 2425
(1990).
\bibitem{gl:apl} \label{gl:apl} S. M. Goodnick and P. Lugli, Appl.
Phys. Lett. {\bf 51}, 584 (1987).
\bibitem{goodnick:lugli} S. M. Goodnick and P. Lugli, in {\it Hot
Carriers in Semiconductor Nanostructures}, edited by J. Shah
(Academic, New York, 1992), p~191.
\bibitem{mosk:thesis} A. Mo\v skov\'a, Ph.D. thesis, Comenius
University, Bratislava (1992).
\bibitem{snoke} D. W. Snoke, W. W. R\"uhle, Y.-C. Lu, and E. Bauser,
Phys. Rev. Lett. {\bf 68}, 990 (1992).
\bibitem{preisel} \label{preisel} M. Preisel, J. M\o rk, and H. Haug,
Phys. Rev. {\bf B~49}, 14~478 (1994).
\bibitem{yokoyama:hess} K. Yokoyama and K. Hess, Phys. Rev.
{\bf B~33}, 5595 (1986).
\end{thebibliography}
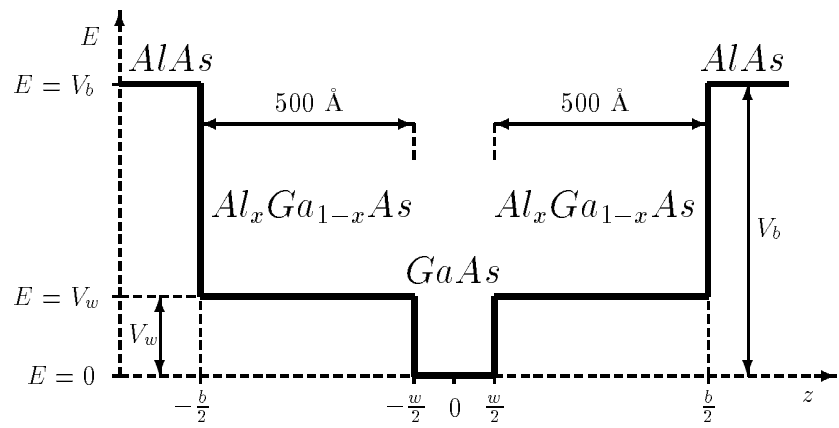
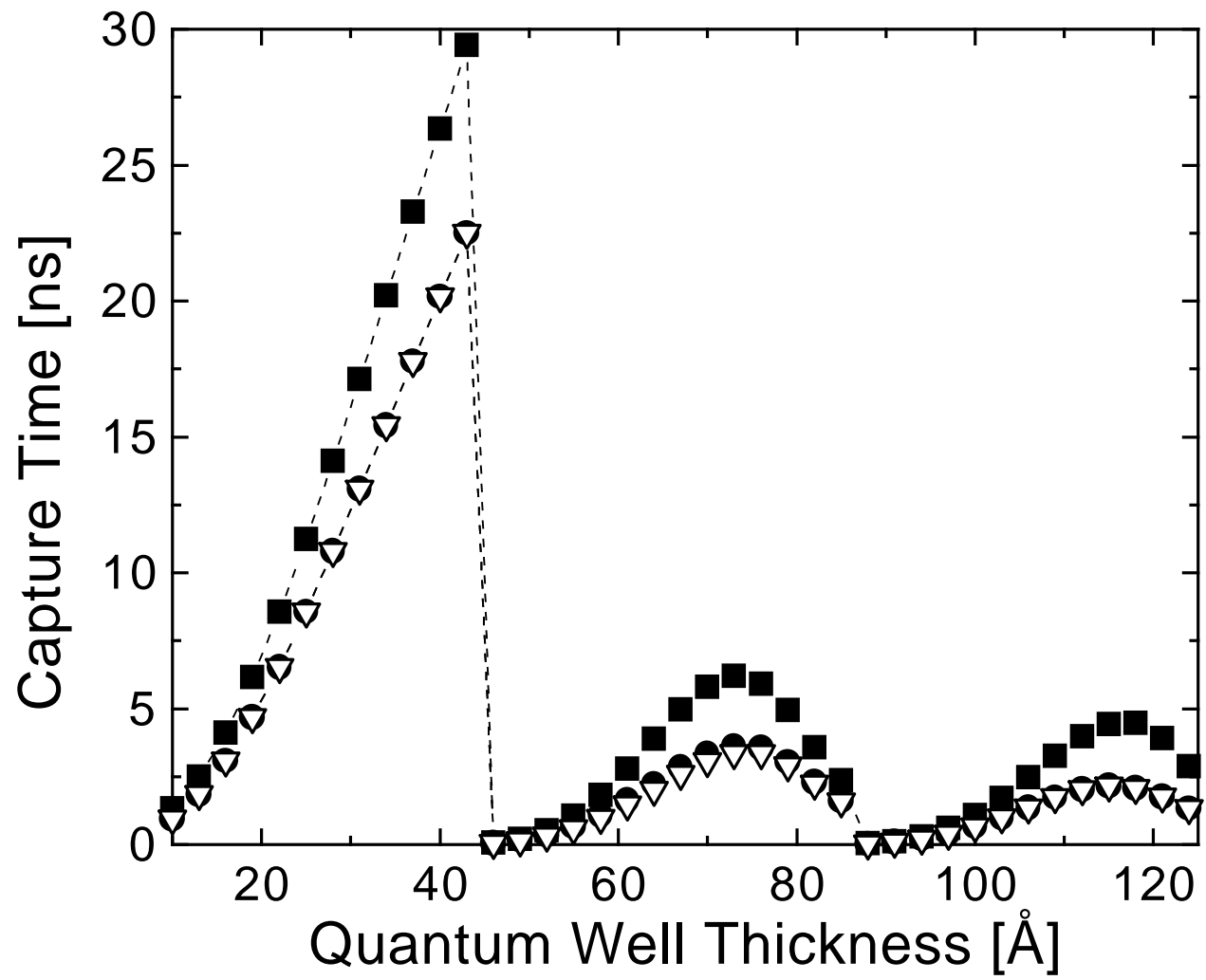
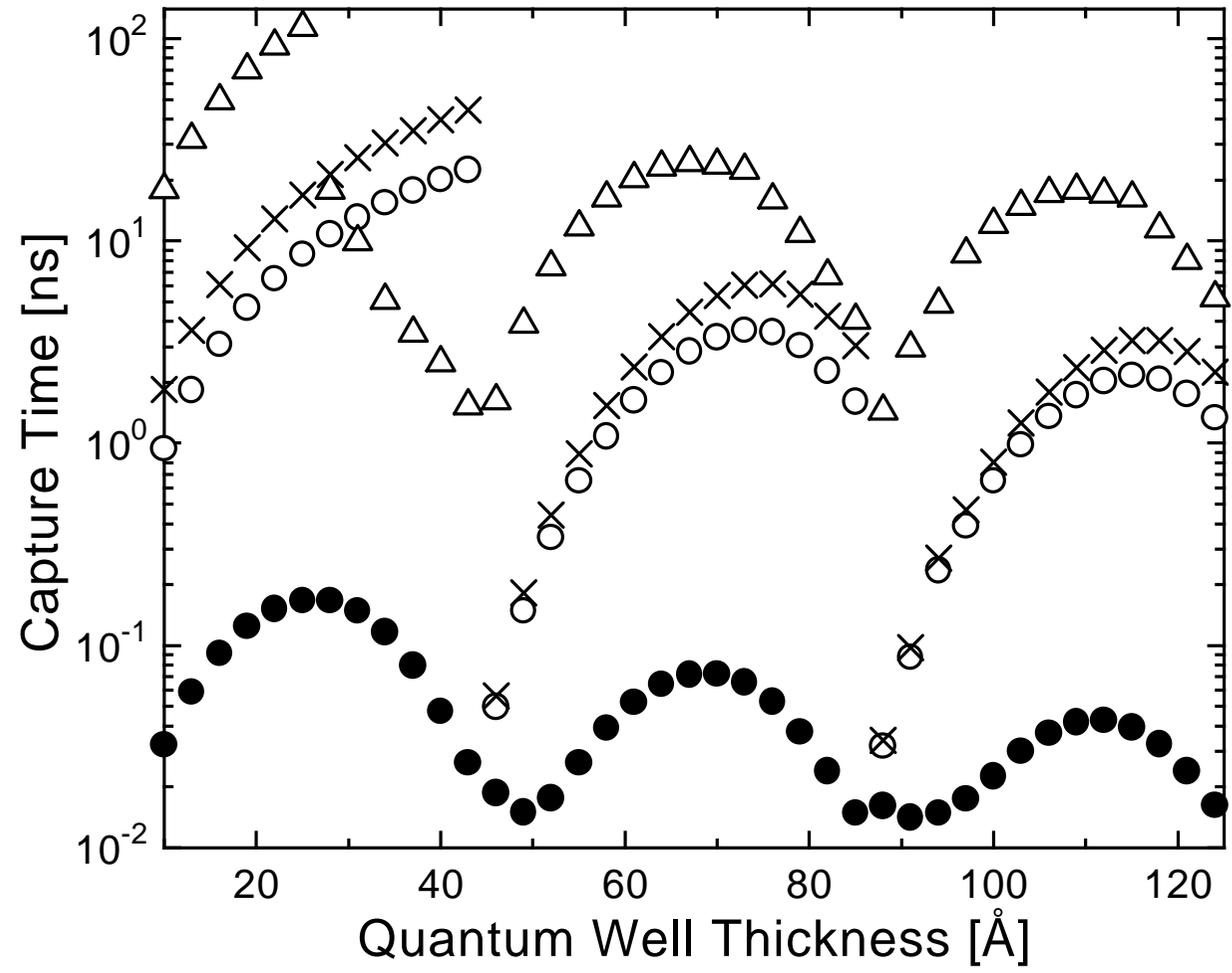
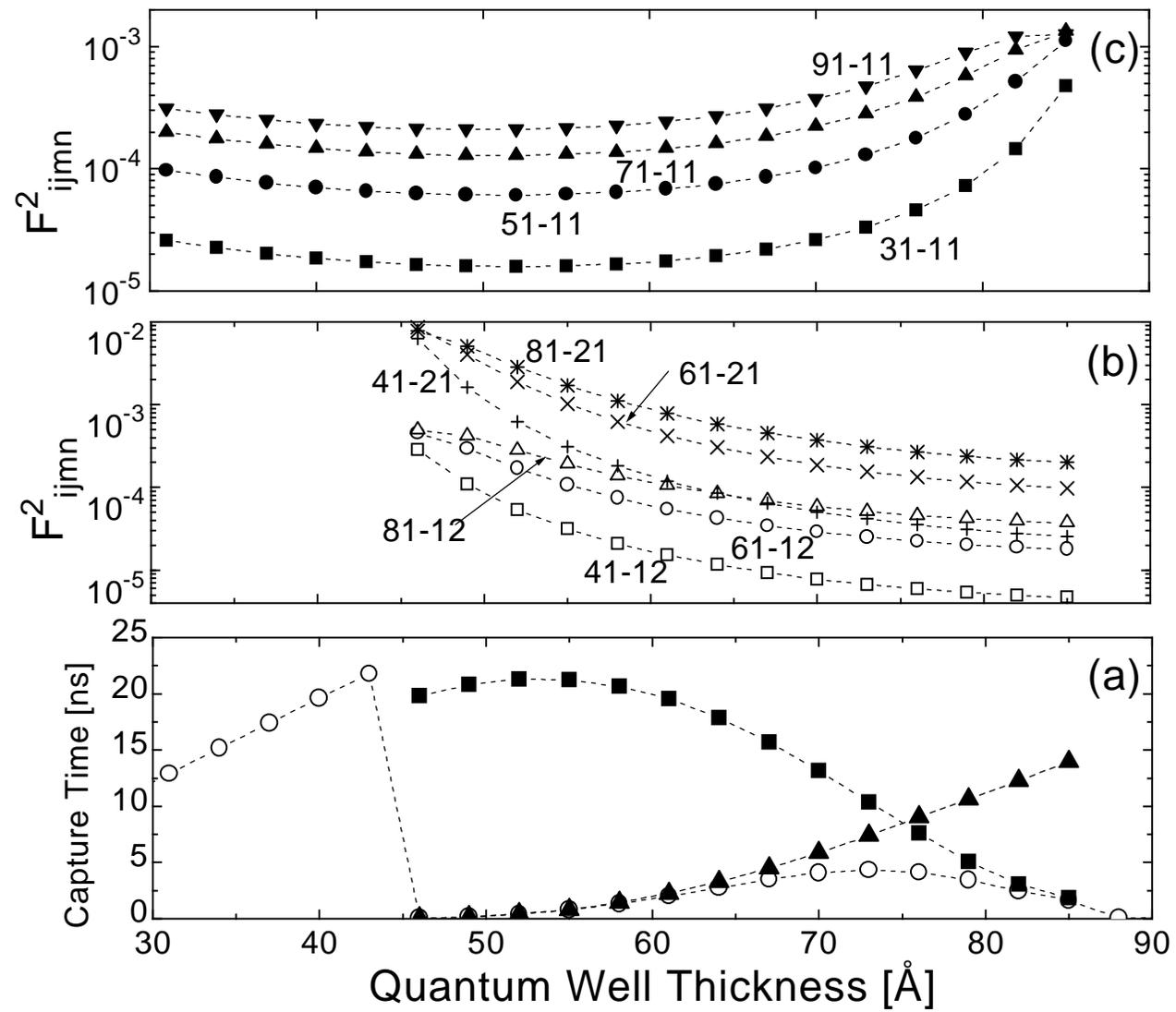
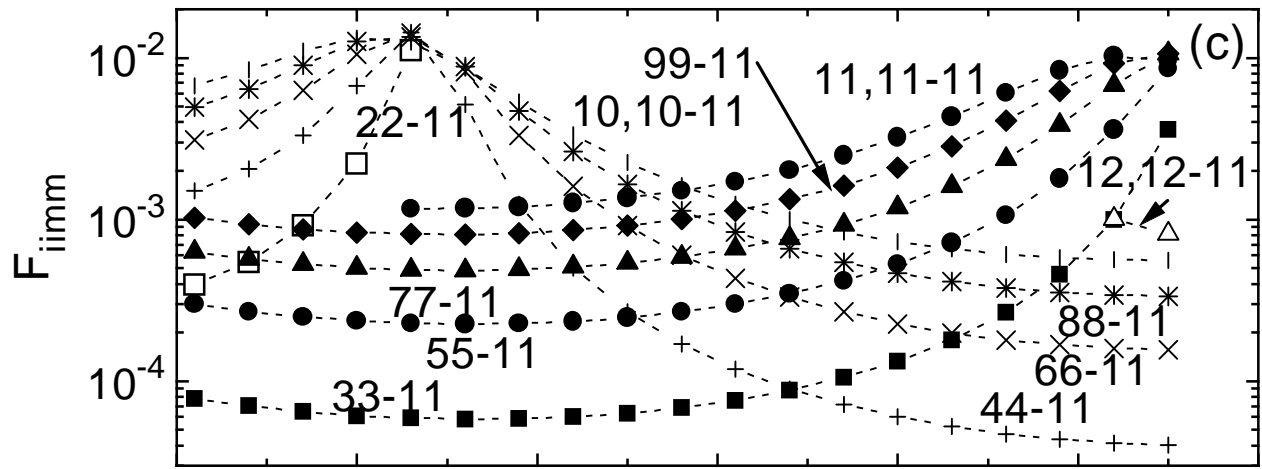
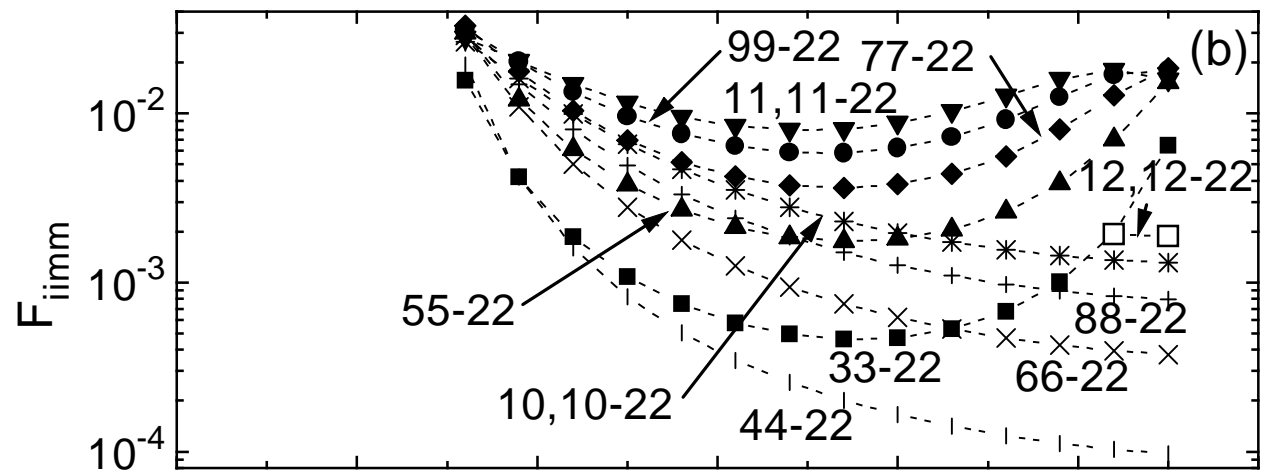
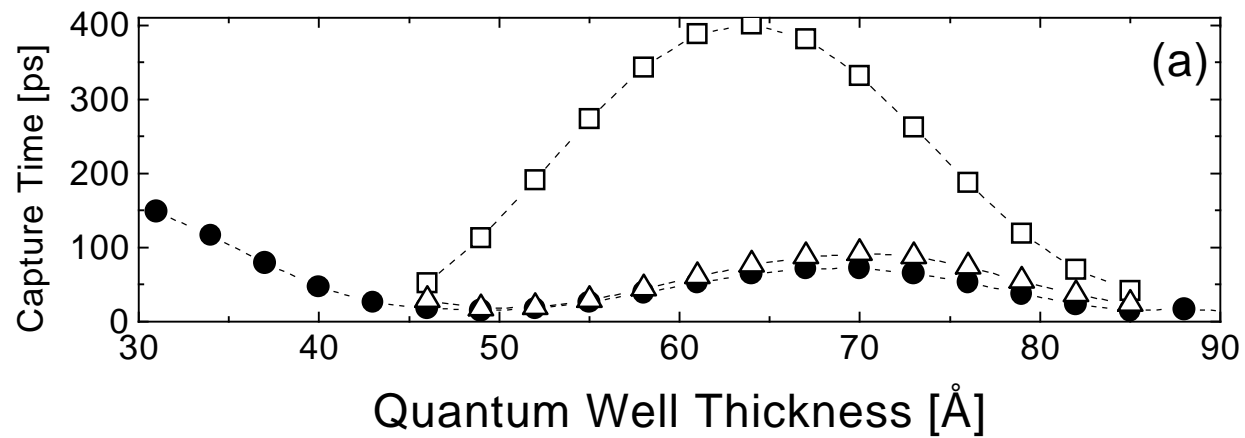
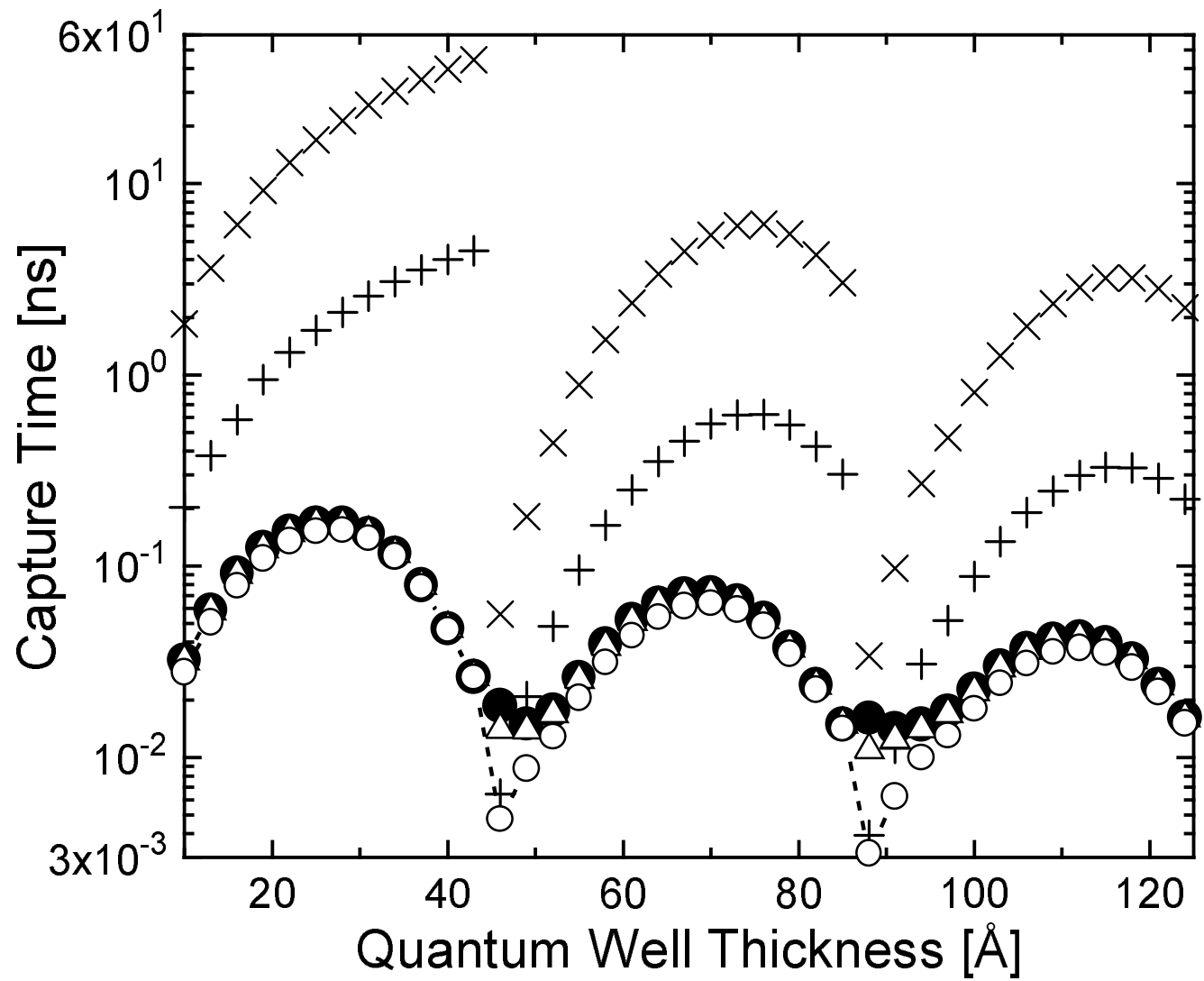
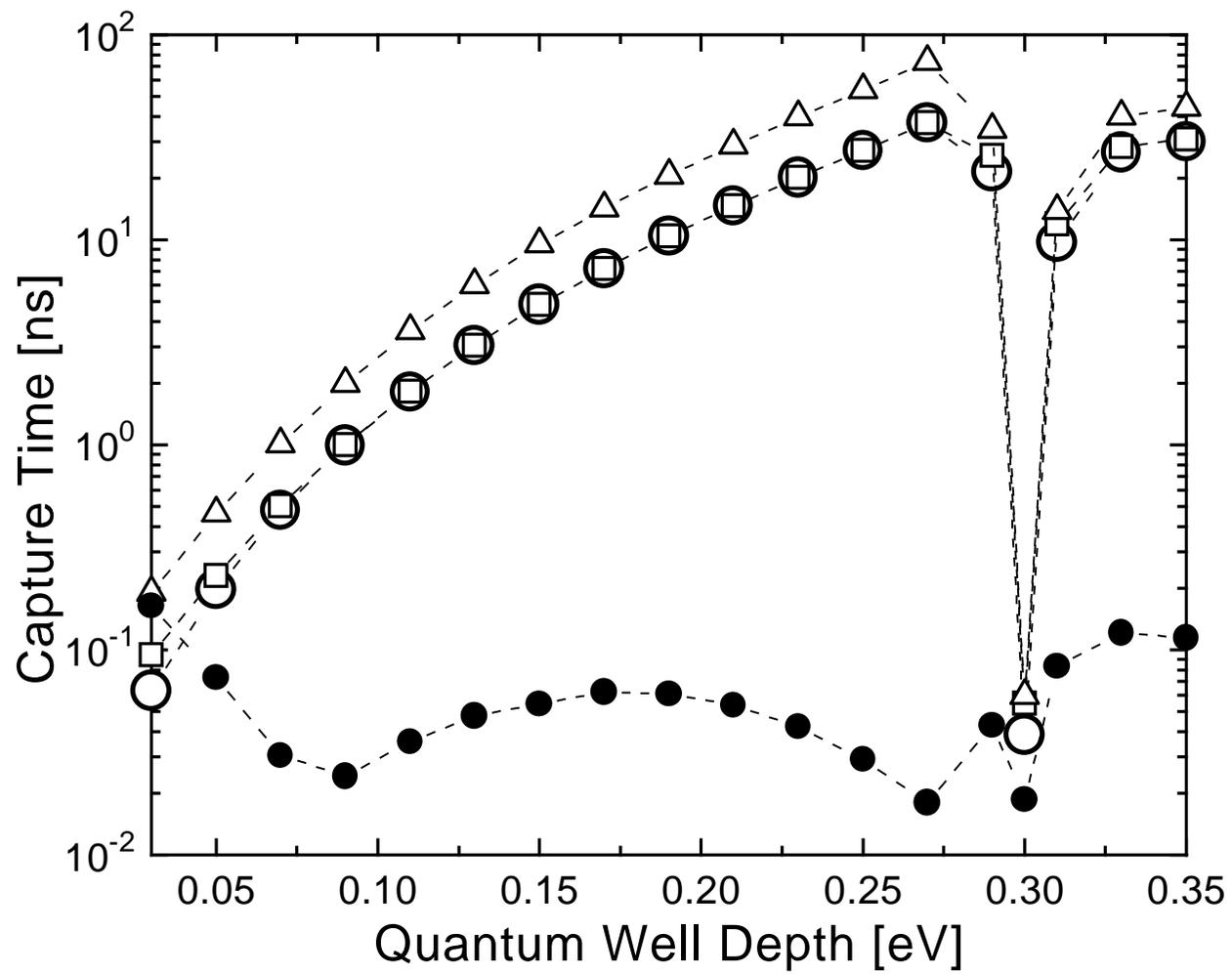
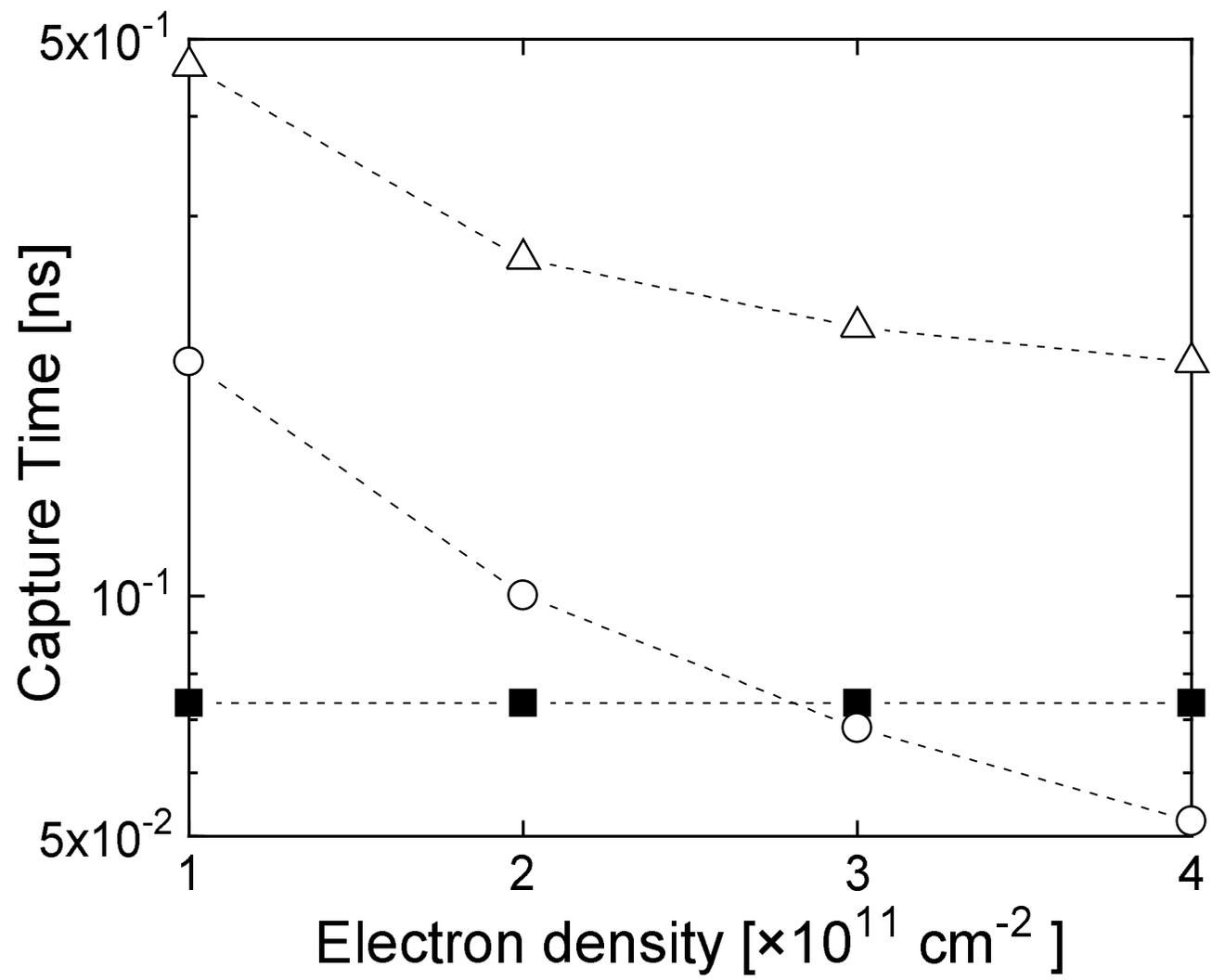
\end{document}